\documentstyle[fleqn,aip2col,epsf]{article}
\let\addressmark\relax

\newcommand{\lsim}{\mathrel{\raisebox{-.6ex}{$\stackrel{\textstyle<}{\sim}$}}}
\newcommand{\gsim}{\mathrel{\raisebox{-.6ex}{$\stackrel{\textstyle>}{\sim}$}}}

\begin{document}
\pagestyle{plain}
\thispagestyle{empty}

\title{\vglue-.6in
\font\fortssbx=cmssbx10 scaled \magstep1
\hbox to \hsize{
\hbox{\fortssbx University of Wisconsin - Madison}
\hfill$\vcenter{\normalsize\hbox{\bf MADPH-97-1011}
                \hbox{August 1997}
                \hbox{\hfil}}$ }\smallskip
Supersymmetry Phenomenology\footnotemark}
\author{\unskip\smallskip
V. Barger\address{Physics Department, University of Wisconsin, Madison, WI 53706, USA}}

\begin{abstract}
The phenomenological implications of a low-energy supersymmetry are surveyed, with particular attention given to unification constraints and the role of a large top quark Yukawa couplings. Generic expectations for sparticle mass spectra are presented along with prospects for their discovery and study at present and future colliders.
\end{abstract}

\maketitle

\renewcommand{\thefootnote}{\fnsymbol{footnote}}
\footnotetext{Talk present at the {\it FCP\,97 Workshop on Fundamental Particles and Interaction}, Vanderbilt University, May 1997.}

\section{Introduction}

The Standard Model (SM) is a pillar of success as an effective theory. Precision experiments agree with SM radiative corrections to an accuracy $\lsim 0.1\%$. However, the SM Higgs sector is problematic. Longitudinal $W$-boson scattering, $W_LW_L\to W_LW_L$, violates unitarity if the Higgs mass exceed about 1~TeV, but the quadratic divergences in radiative corrections to the Higgs mass give a Higgs mass that is naturally of the order of the Planck mass. A way out of this conundrum is a low-energy fermion-boson supersymmetry (SUSY) in which each SM fermion (boson) has a boson (fermion) superpartner; see Table~1. Two Higgs doublets are required in the minimal supersymmetric standard model (MSSM), one ($ H_u$) to give mass to the up-type quarks and leptons and the other ($H_d$) to give mass to the down-type fermions. The $\tilde W^\pm, \tilde H^\pm$ sparticle states mix and their spin-1/2 mass eigenstates are the charginos, denoted by $\chi_{1,2}^\pm$.  Similarly, the neutral sparticles states $\tilde B^0, \tilde W^0, \tilde H^0_u, \tilde H^0_d$ mix to give the neutralino mass eigenstates $\chi^0_{1,2,3,4}$ In exact SUSY, a particle and its sparticle companion have the same mass and couplings. In broken SUSY the sparticles have higher masses than the particles but the exact SUSY coupling relationships are maintained. The additional radiative contributions to the Higgs mass from sparticle loops cancel the quadratic divergence SM loop contribution, solving the naturalness and gauge hierarchy problems.
There is a vast literature on supersymmetry phenomenology and the reader may consult recent reviews\cite{reviews} and textbooks\cite{books,HHG} for references.

\begin{table}[t]
\caption{}
\centering
\leavevmode
\epsfxsize=3.2in\epsffile{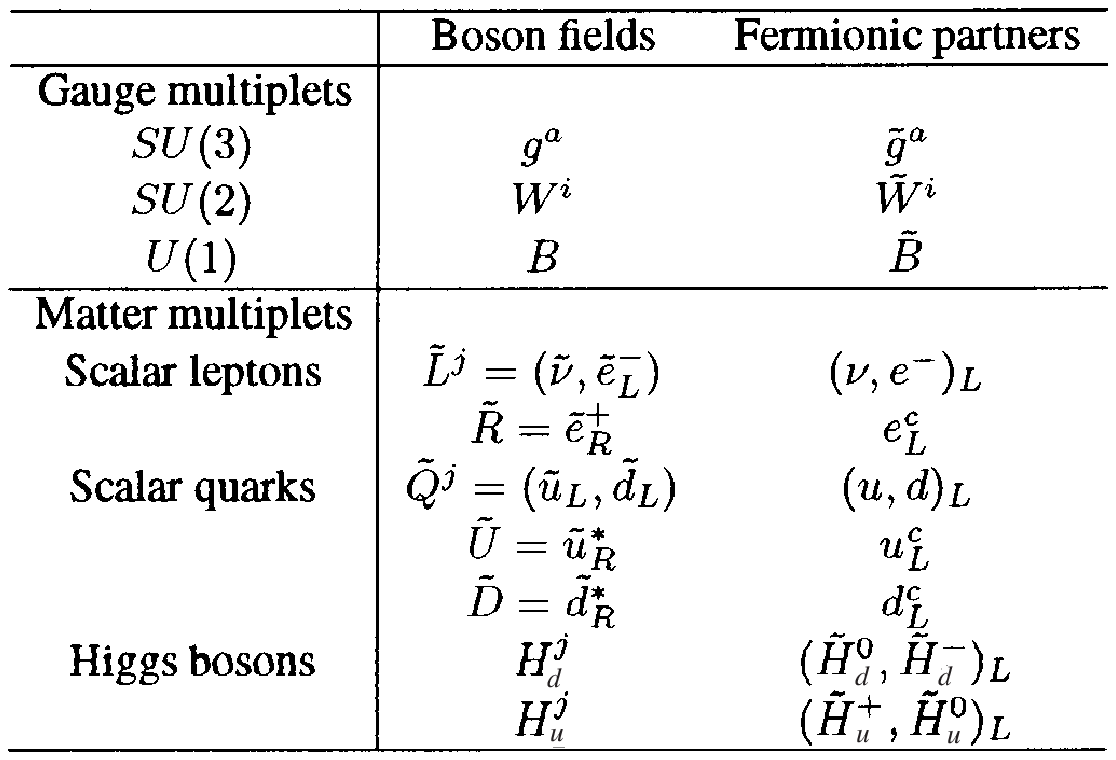}
\end{table}

\section{Gauge Coupling Unification}

The Renormalization Group Equations (RGE), found by analyzing loop corrections, predict the evolution of the couplings with energy scale. The $\alpha_i =  g_i^2/(4\pi)$, with label $i=1,2,3$ for U(1), SU(2), SU(3) satisfy the RGEs,
\begin{equation}
{d\alpha_i\over dt} = {1\over2} \left[ b_i \alpha_i^2 + {1\over4\pi} \sum_{j=1}^3 b_{ij} \alpha_i^2 \alpha_j + \cdots \right]
\end{equation}
where $t = \ln(\mu/M_G)$ and $\mu$ is the running mass parameter. The $b_i$ are known constants from the particle content of the loops. The running of the strong coupling constant $\alpha_s(t)\equiv\alpha_3$ is now convincingly established; see Fig.~1\cite{marti}.

\begin{figure}[t]
\centering
\leavevmode
\epsfxsize=3.2in\epsffile{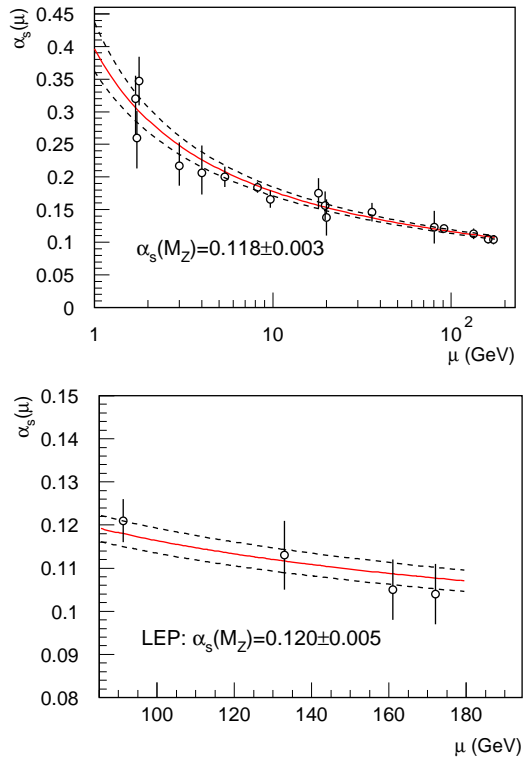}

\caption[]{Running of the QCD coupling constant (from Ref.~\cite{marti}).}
\end{figure}

Starting from the measured $\alpha_i$ at $\mu=M_Z$, the couplings can be extrapolated to higher scales, assuming that only particles of low mass contribute to the loops (i.e., there is a particle desert between the TeV scale and the unification scale). If a Grand Unified Theory (GUT) exists at a high scale $M_G$, then the three couplings should evolve to a common point of intersection. Unification of the gauge couplings is remarkably realized if supersymmetric particle masses are between $\sim100$~GeV and $\sim1$~TeV, as illustrated in Fig.~2 for a 1~TeV SUSY mass scale. Unification occurs for $\alpha_s(M_Z)=0.13\pm0.01$ (see e.g.\ Ref.~\cite{langetal}), where the uncertainty is associated with the SUSY mass spectrum. This value is remarkably consistent with the LEP measurement $\alpha_s(M_Z) = 0.120\pm 0.005$\cite{marti}. No such common intersection of the couplings occurs in the SM.

\begin{figure}
\centering
\leavevmode
\epsfxsize=3in\epsffile{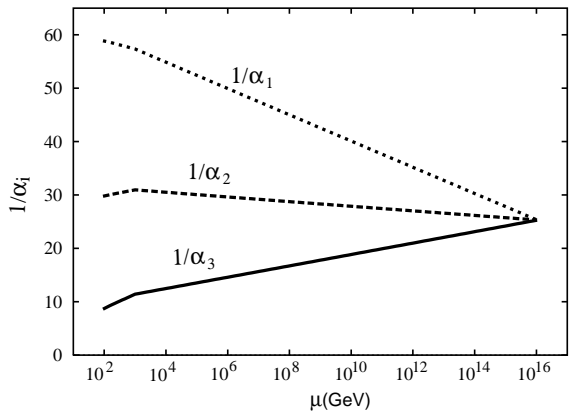}

\caption[]{Gauge coupling evolution in the MSSM with a SUSY scale of 1~TeV (from \cite{BBO}).}
\end{figure}

\section{Yukawa Coupling Evolution}

The fermion masses are generated by the vacuum expectation values (vevs) $v_u$ of the $H_u^0$ and $v_d$ of the $H_d^0$, which are expressed in terms of the SM vev as
\begin{equation}
v_u = v\sin\beta \,,\qquad v_d = v\cos\beta \,.
\end{equation}
Then the third generation masses are related to the Higgs Yukawa couplings $\lambda_f$ as
\begin{equation}
\begin{array}{c}
m_t(m_t) = \lambda_t(m_t) v_u\,, \quad
m_b(m_b) = \lambda_b(m_b) v_d\,, \\ \noalign{\vskip1ex}
m_\tau(m_\tau) =\lambda_\tau(m_\tau) v_d \,.
\end{array}
\end{equation}
The Yukawa couplings also evolve with scale. In terms of the $Y_f \equiv \lambda_f^2/(4\pi)$, the RGEs for the Yukawa couplings at one-loop order are
\begin{eqnarray}
{dY_t\over dt} &=& {1\over2\pi} Y_t \left( 6Y_t + Y_b - {16\over3}\alpha_3 - 3\alpha^2 \right) \,, \\
{dY_b\over dt} &=& {1\over2\pi} Y_b \left( Y_t + 6Y_b  + Y_\tau - {16\over3\alpha_3} - 3\alpha_2 \right) \, \\
{dY_\tau\over dt} &=& {1\over2\pi} Y_\tau \left( 3Y_b + 4Y_\tau + 3\alpha_2 \right) \,.
\end{eqnarray}
Here the small contributions from $\alpha_1$ have been ignored. To correctly predict $m_b/m_\tau$ with $Y_b= Y_t$ unification\cite{btau}, large values of $Y_t$ (close to the perturbative bound) at the GUT scale are required\cite{BBO}. In this circumstance evolution drives $\lambda_t$ towards a quasi-infrared fixed point ($d\lambda_t/dt\simeq0$)\cite{BBO,IRFP1,IRFP2,IRFP3} independent of the precise value of $\lambda_t$ at the GUT scale.

Figure 3 shows the $\lambda_t$ infrared fixed point band in the $\tan\beta$ vs.\ $m_t$ plane. For $m_t = 175$~GeV there is a low $\tan\beta$ solution with $\tan\beta\simeq1.8\ (\beta\simeq60^\circ)$, given by
\begin{equation}
m_t = 200\rm\ GeV \sin\beta \,.
\end{equation}
There is also a large $\tan\beta$ fixed point solution with $\tan\beta\simeq56$ for which triple Yukawa coupling unification ($\lambda_t = \lambda_b = \lambda_\tau$) is approximately realized at the GUT scale\cite{IRFP2}. These fixed point solutions are very attractive theoretically, though it is too soon to rule out values of $\tan\beta$ between the two fixed points, or even below.

\begin{figure}[t]
\centering
\leavevmode
\epsfxsize=3.2in\epsffile{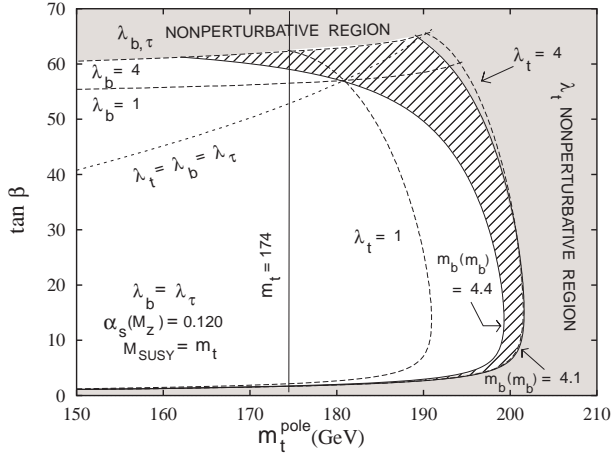}

\caption[]{Contours of fixed $b$-quark mass in the plane of $\tan\beta$ versus the top-quark mass, along with contours of constant GUT scale Yukawa couplings. The top Yukawa infrared fixed point region is given by the line shading. From Ref.~\cite{BBO}.}
\end{figure}

\section{Soft SUSY Breaking}

Supersymmetry is usually presumed to be broken in a hidden sector with this breaking transmitted to the observable sector  via a)~gravitational interactions ($N=1$ supergravity) or b)~gauge sector interactions of a messenger sector with effective scale $\lsim 100$~TeV. Both categories of model contain ``soft" breaking mass terms that do not reintroduce quadratic divergences. All gauge-invariant soft mass terms are included in the Lagrangian
\begin{eqnarray}
{\cal L}_{\rm soft} &=& -\sum_{\rm scalars} m^2 A^2 
 - \sum_{\rm gauginos} M(\lambda\lambda + \bar\lambda\bar\lambda)\nonumber\\
&& {}  + B\mu\epsilon_{ij} \hat H_1^i \hat H_2^j+ A_t \lambda_t \hat Q\hat U \hat H_u\nonumber\\
&& {} + A_b\lambda_b \hat Q\hat D \hat H_d + A_\tau \lambda_\tau \hat L \hat E \hat H_d \,.
\end{eqnarray}
Subsequently we concentrate on the phenomenology of the minimal supergravity model (mSUGRA), which assumes universal soft SUSY breaking parameters at the GUT scale:
\begin{flushleft}
\begin{tabular}{ll}
$m_0$ & common scalar mass\\
$m_{1/2}$ & common gaugino mass\\
$A_0$ & common trilinear coupling\\
$B_0$ & common bilinear coupling\\
$\mu_0$ & Higgs mixing mass
\end{tabular}
\end{flushleft}
Starting from these universal parameters at the GUT scale, the soft parameters at the electroweak scale are obtained from RGE evolution. Figure~4 shows typical results of the evolution. The Higgs miracle is explained by the evolution: With a large $\lambda_t$ at the GUT scale the mass-squared of $H_u$ is driven negative at the electroweak scale\cite{A_t}. The values of $|\mu(M_Z)|$ and $B(M_Z)$ are then determined by the minimization of the Higgs potential. At tree level the minimization conditions are:
\begin{eqnarray}
\mu^2 &=& {m_{H_d}^2 - m_{H_u}^2 \tan^2\beta\over\tan^2\beta-1} - {1\over2} M_Z^2 \,,\\
B\mu &=& -{1\over2}\left( m_{h_d}^2 + m_{H_u}^2 + 2\mu^2 \right) \sin^2\beta\,.
\end{eqnarray}
For the $\lambda_t$ fixed point solution at low $\tan\beta$ the $A_t$ parameter also approaches a fixed point, independent of its GUT scale value\cite{IRFP3}. Thus the low-energy phenomenology of the mSUGRA models is given in terms of the parameters $m_0$, $m_{1/2}$, sign of $\mu$, and $\tan\beta$.

\begin{figure}
\centering
\leavevmode
\epsfxsize=3.4in\epsffile{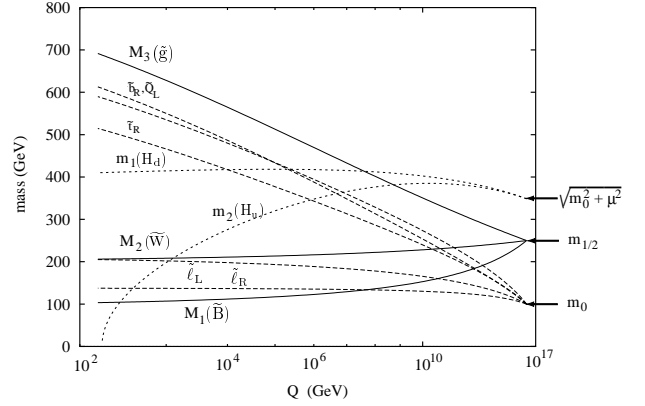}

\caption[]{Evolution of the sparticle spectrum down to the electroweak scale from universal boundary conditions at the GUT scale. From Ref.~\cite{BBO}.}
\end{figure}

The evolution of the gaugino masses $M_1$, $M_2$, $M_3$ depends only on the gauge couplings
\begin{equation}
{dM_i\over dt} = -b_i\alpha_i M_i\,,
\end{equation}
where $b_1=33/5$, $b_2=1$, $b_3=-3$,
so the $M_i$ at any scale are simply related,
\begin{equation}
M_3/\alpha_3 = M_2/\alpha_2 = M_1/\alpha_1 \,.
\end{equation}
The values of the $M_i$ at the $M_Z$ scale are given in terms of the GUT scale gaugino mass by
\begin{equation}
M_3 \simeq 3.2 m_{1/2}\,,\ 
M_2 \simeq 0.88 m_{1/2}\,,\ 
M_1 \simeq 0.44 m_{1/2}\,.
\end{equation}
The gluino mass is $M_3$. In the following we discuss the qualitative features of sparticle masses for the low $\tan\beta$ $\lambda_t$ fixed point, for which $|\mu| \gg M_Z$ is obtained from RGE evolution.

\section{Chargino Sector}

In a first approximation the mass matrix in the $(\tilde W^\pm, \tilde H^\pm)$ basis is diagonal,
\begin{equation}
{\cal M} = \left( \begin{array}{cc} M_2& 0\\ 0& -\mu \end{array} \right)\,.
\end{equation}
Thus the chargino mass eigenstates are
\begin{equation}
\chi_1^\pm \sim \tilde W^\pm\,,\quad \tilde\chi_2^\pm \sim H^\pm \,.
\end{equation}

\section{Neutralino Sector}

Here the approximate mass matrix in the $(\tilde B^0, \tilde W^0, \tilde H_u, \tilde H_u^0)$ basis is
\begin{equation}
{\cal M} = \left( \begin{array}{cccc}
M_1& 0 & & \\
0& M_2& & \\
& & 0& \mu\\
& & \mu& 0
\end{array} \right)
 \,.
\end{equation}
Thus the two lightest neutralinos are approximately gauginos,
\begin{equation}
\chi_1^0 \sim \tilde B^0\,, \quad \chi_2^0 \sim \tilde W^0 \,.
\end{equation}
The masses of the lightest color singlet ino states and the gluino are approximately related as
\begin{equation}
\chi_1^0 : \chi_2^0 : \chi_1^\pm : \tilde g = 1:2:2:7 \,.
\end{equation}

\section{Stop Sector}

The stop mass-squared matrix in the $\tilde t_L, \tilde t_R$ basis is
\begin{equation}
m^2 = \left( \begin{array}{cc}
L^2 & am_t\\
am_t & R^2
\end{array} \right) \,,
\end{equation}
where
\begin{eqnarray}
a &=& A_t + \mu\cot\beta \,,\\
L^2 &=& m_t^2 +M_{\tilde Q}^2 - 0.35M_Z^2 |\cos2\beta| \,,\\
R^2 &=& m_t^2 + M_{\tilde U}^2 - 0.15 M_Z^2 |\cos2\beta| \,.
\end{eqnarray}
The off-diagonal terms are proportional to $m_t$. Consequently there may be large mixing. Diagonalization of $m^2$ leads to two stop mass eigenstates $\tilde t_1, \tilde t_2$. The $\tilde t_1$ may be light if the mixing is large or if $M_{\tilde U}^2 \lsim 0$. The stop masses and mixings determine the precise value of the light Higgs boson mass through the radiative corrections to $m_h$. For example, in the limit of large mass of the CP-odd Higgs state $A$, the mass of the lightest CP-even Higgs boson $h$ is given by\cite{higgs-lep2}
\begin{eqnarray}
m_h^2 &=& M_Z^2 \cos2\beta \left( 1 - {3m_t^2 t\over 8\pi^2 v^2}\right) \nonumber\\
&& \hspace{-.6in} {}+ {3m_t^4\over 4\pi^2v^2} \left[ t + {\tilde\kappa\over 2} + \left( {3\over 32\pi^2} {m_t^2\over v^2} - {2\alpha_s\over\pi} \right) \left(\tilde\kappa t + t^2\right) \right] ,
\end{eqnarray}
with
\begin{eqnarray*}
t &=& \ln \left( M_S^2\over m_t^2 \right) \\
\tilde\kappa &=& {2\tilde A_t\over M_S^2} \left( 1 - {\tilde A_t^2\over 12M_S^2}\right) \\
\tilde A_t &=&  A_t + \mu\cot\beta \,,
\end{eqnarray*}
where $M_S^2 = (m_{\tilde t_1}^2 + m_{\tilde t_2}^2)/2$. The radiative corrections can substantially increase the tree-level bound $m_{h^0} \leq M_Z |\cos2\beta|$ to $m_{h^0} \lsim 130$~GeV.

\section{Generic SUSY Mass Spectra}

Representative results for the SUSY mass spectra for the low $\tan\beta$ fixed point scenario are shown in Fig.~5, where the predicted masses for $m_{1/2} = 150$~GeV are given versus $m_0$. Typical superpartner masses are
\begin{flushleft}
\begin{tabular}{ll}
selectron & $\gsim70$ GeV\\
sneutrino & $\gsim100$ GeV\\
stop & $\gsim90$ GeV\\
chargino & $\gsim100$ GeV\\
LSP & $\gsim50$ GeV\\
gluino & $\gsim350$ GeV
\end{tabular}
\end{flushleft}
The production and decays of the sparticles offer many interesting possibilities for experimental searches.

\begin{figure}[t]
\centering
\leavevmode
\epsfxsize=2.5in\epsffile{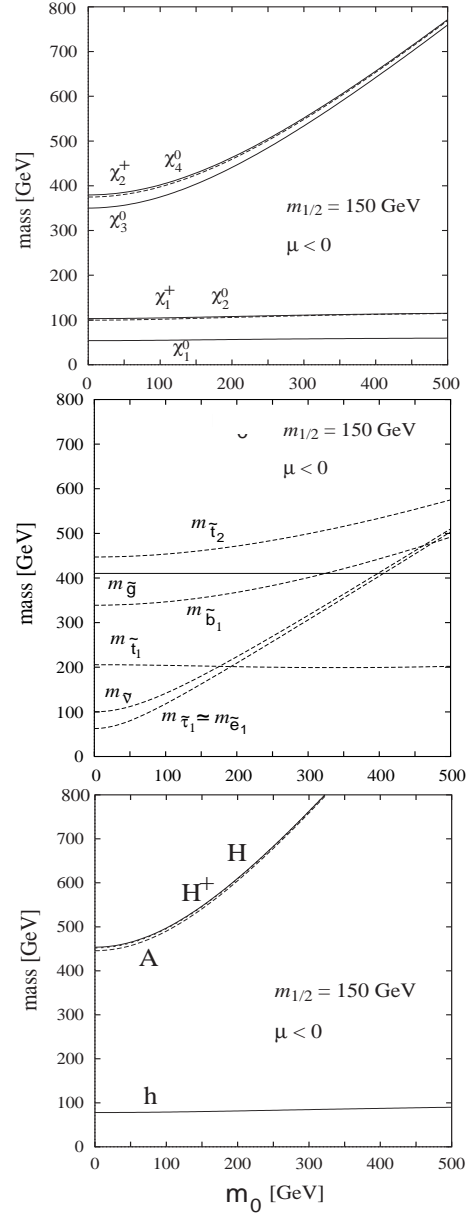}

\caption[]{Characteristic mass spectra for sparticles in the mSUGRA model (from Ref.~\cite{BBO}.)}
\end{figure}

\section{Light Higgs Search}

Ongoing searches for Higgs bosons at LEP-2 are based on the processes
\begin{eqnarray}
&& e^+e^-\to (h\mbox{ or }H)Z \to b\bar b q\bar q, b\bar b\ell\bar\ell, b\bar b\nu\bar\nu, \tau\bar\tau q\bar q \,,\\
&& e^+e^-\to hA \to b\bar b b\bar b, \tau\bar\tau b\bar b\,.
\end{eqnarray}
Figure 6 shows the regions of the $(\tan\beta, m_h)$ plane presently excluded ($M_S = 1$~TeV is assumed)\cite{aleph}.

\begin{figure}
\centering
\leavevmode
\epsfxsize=3.2in\epsffile{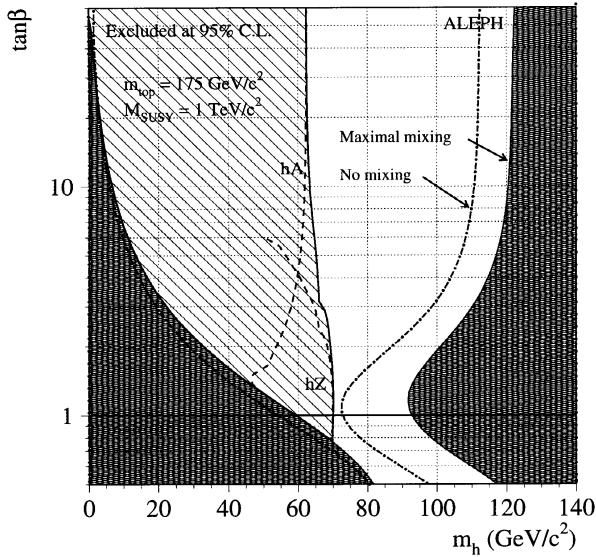}

\caption{Excluded regions of the MSSM parameter space. (The dark areas are theoretically disallowed; this region depends on the extent of stop mixing.) The hatched area is excluded at 95\% confidence level by the combined search for $e^+e^-\to hZ$ and $e^+e^-\to hA$. From Ref.~\protect\cite{aleph}}
\end{figure}

Upgrades at the Tevatron collider (Main Injector, TeV-33) will allow for a SM Higgs search up to $\sim$120~GeV and possibly higher through the process\cite{stange}
\begin{equation}
q\bar q\to Wh, \quad q\bar q\to t\bar th
\end{equation}
using vertex detectors to identify $h\to b\bar b$ decays. If $m_A$ is large, then the couplings of the lightest MSSM Higgs boson are essentially the same as the SM Higgs, and the mass reach for the MSSM Higgs is comparable to that above for the SM Higgs.

At the LHC at least one of the MSSM Higgs bosons should be found in any region of the $(\tan\beta, m_A)$ parameter space\cite{froid}. For the lightest MSSM Higgs, the decays $h\to \gamma\gamma$ and $h\to ZZ^*\to 4\,$leptons ($\ell=e,\mu$) are important search modes, with the $h$ produced by gluon-gluon fusion ($gg\to h$).

\section{Neutralino Dark Matter}

A discrete quantum number known as $R$-parity, $R=(-)^{3B+L+2S}$, is commonly introduced in supersymmetry models to keep the proton sufficiently stable: $R=+1$ for particles and $R=-1$ for sparticles. Then sparticles are produced only in pairs and the lightest supersymmetric particle (LSP) is stable. In mSUGRA models the LSP is the lightest neutralino, $\chi_1^0$. 

In the early Universe, when the temperature exceeded $m_{\chi_1^0}$, LSPs would have existed abundantly in thermal equilibrium, with the annihilation rate balanced by pair production. However, after the temperature dropped below $m_{\chi_1^0}$ and the annihilation rate dropped below the expansion rate of the Universe, a relic cosmological abundance of the LSP would remain; the LSP is an attractive candidate for the observed dark matter\cite{jungman}. In terms of $\Omega\equiv\rho/\rho_c$, where $\rho_c$ is the critical mass density to close the universe, clusters and large scale structure indicate $\Omega>0.2$. However, the nucleosynthesis of the heavy elements constrains $\Omega_b h^2 \leq 0.03$, where $h$ is the Hubble constant in units of 100~km$^2$/s/Mpc. Recent determinations find $h\simeq 0.65$. Thus the cosmologically interesting region for neutralino dark matter is
\begin{equation}
0.1\leq \Omega_{\chi_1^0} h^2 < 0.5 \,.
\end{equation}
Figure 7 illustrates regions of the mSUGRA $m_0$ and $m_{1/2}$ parameters that are compatible with the cosmological constraint, for the low $\tan\beta$ fixed point solution\cite{BK}. The neutralino explanation of dark matter indicates a low mass scale for supersymmetry in this fixed point scenario.

\begin{figure*}[t]
\centering
\leavevmode
\epsfxsize=5in\epsffile{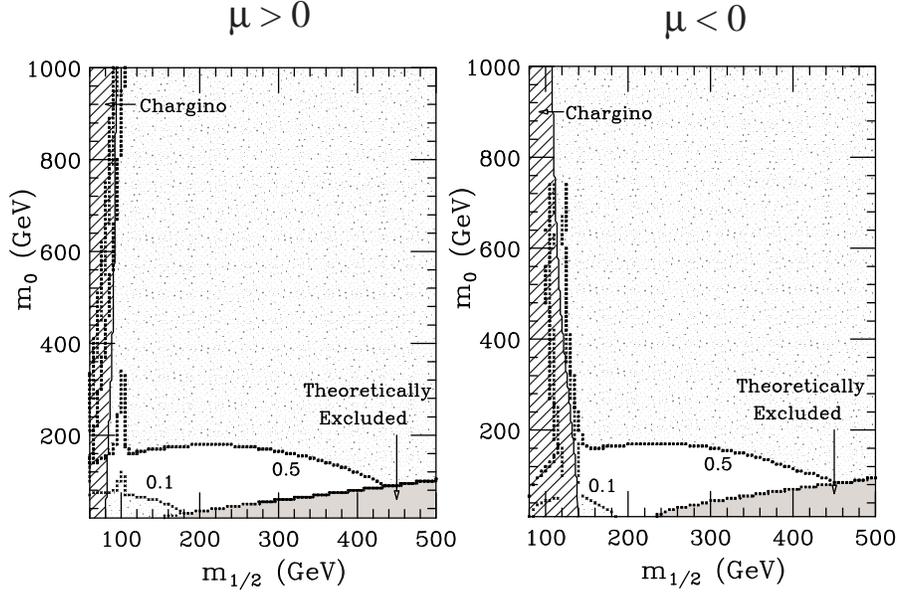}
\vspace{-.5in}

\caption[]{Contours of neutralino relic density $\Omega_{\chi_1^0}h^2 = 0.1$ and 0.5 in the $(m_{1/2}, m_0)$ plane for $\tan\beta=1.8$. The shaded regions are excluded by (i)~a cosmologically uninteresting relic density, (ii)~theoretical requirements, or (iii)~the chargino search at LEP2. From Ref.~\cite{BK}.}
\end{figure*}

\section{Finding Sparticles at Colliders}

At $e^+e^-$ or $\mu^+\mu^-$ colliders, the production of sparticles may give very clean signatures. Two such examples are
\begin{eqnarray}
&&\hspace{-.2in} e^+e^-\to \chi_1^+\chi_1^-\to (W^+\chi_1^0) (W^-\chi_1^0) \,,\\
&&\hspace{-.2in} e^+e^-\to \tilde\mu^+\tilde\mu^-\to (\mu^+\chi_1^0) (\mu^-\chi_1^0)\,.
\end{eqnarray}
The $W$-bosons from the chargino decay can be real or virtual. From LEP-2 searches at c.m.\ energies $\sqrt s = 161$--172~GeV, the ALEPH collaboration has placed the limits\cite{aleph2}
\begin{equation}
M_{\chi_1^\pm} > 85{\rm\ GeV},\ 
M_{\tilde\mu^\pm} > 70{\rm\ GeV},\ 
M_{\tilde t_1} > 63{\rm\ GeV}.
\end{equation}

Post discovery, the next step will be to determine the sparticle spins, masses and couplings. The two-body decay $\tilde\mu_R\to\chi_1^0\mu^-$ is a particularly simple example. The two endpoints of the flat energy spectrum for the decay muon give the relations
\begin{eqnarray}
{m_{\tilde\mu}^2 - m_{\chi_1^0}^2\over 2m_{\tilde \mu} } &=& 
\sqrt{1-\beta_{\tilde\mu}\over 1+\beta_{\tilde\mu}} \left(E_\mu\right)^{\rm max}_{\rm lab} \nonumber\\
&=& \sqrt{1+\beta_{\tilde\mu}\over 1+\beta_{\tilde\mu}} \left(E_\mu\right)^{\rm min}_{\rm lab} \,,
\end{eqnarray}
where $\beta_{\tilde\mu} = \left( 1 - 4m_{\tilde\mu}^2 /s\right)^{1/2}$ is the smuon velocity. Thus the measured muon energy endpoints determine both $m_{\tilde\mu}$ and $m_{\chi_1^0}$\cite{nlc}. Figure~8 shows the results of a realistic simulation. Similar mass determinations from kinematic endpoints of other decay processes are possible at hadron colliders\cite{paige}.

\begin{figure}[t]
\centering
\leavevmode
\epsfxsize=3in\epsffile{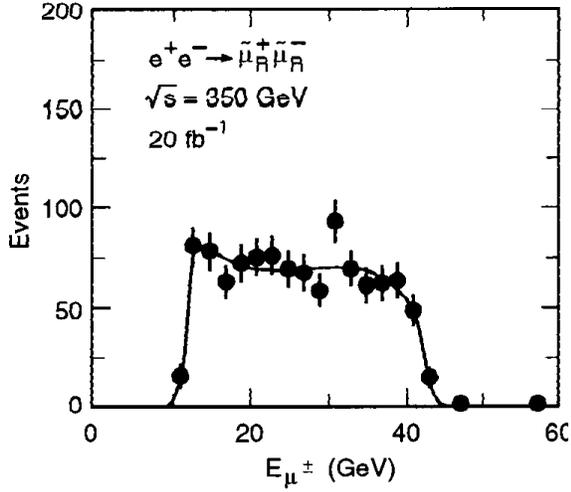}

\caption[]{Simulation of the $\tilde \mu_R$ mass measurement at an $e^+e^-$ linear collider (from Ref.~\cite{tsuk}).}
\end{figure}

Heavy sparticles will decay through multistep cascades. An example is gluino decay through the following chain
\begin{equation}
\tilde g\to q\tilde q\,,\quad \tilde q\to q\tilde\chi_i\,,\quad \chi_i\to\chi_j W\,,\quad \chi_j\to f\bar f\chi_1^0 \,.
\end{equation}
The signatures of pair-produced heavy sparticles are isolated leptons, missing transverse energy (LSPs and neutrinos), and jets. Missing $E_T$ searches at the Tevatron exclude regions of gluino and squark masses as shown in Fig.~9. The trilepton signal from $u\bar d\to W^{+*}\to \chi_1^+\chi_2^0$ with $\chi_1^+\to \chi_1^0\ell^+\nu$ and $\chi_2^0\to\chi_1^0\ell^+\ell^-$ decays $(\ell=e,\mu)$ gives the highest future SUSY mass reach at the Tevatron collider\cite{BT}. At the LHC many SUSY channels are accessible and there is a good safety margin for discovery up to the TeV scale. The comparative reach of the Tevatron, LHC and NLC is illustrated in Fig.~10 in the space of mSUGRA scalar and gaugino masses\cite{susy-wg}. Contours of 1~TeV gluino and squark masses are given in the figure for reference.

\begin{figure}[t]
\centering
\leavevmode
\epsfxsize=3in\epsffile{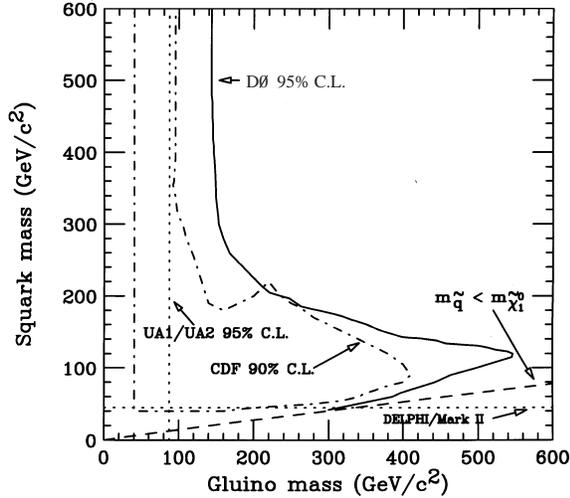}

\caption[]{Squark and gluino mass regions excluded by Tevatron searches (from Ref.~\cite{d0}).}
\end{figure}

\section{Conclusion}

Supersymmetry is a compelling extension of the Standard Model which solves the quadratic divergence problem, gives unification of the gauge couplings, and accounts for the dark matter in the Universe. The heavy top quark plays a pivotal role in unified SUSY models. A large top-quark Yukawa coupling, needed for $b$-$\tau$ unification and to explain electroweak symmetry breaking as a radiative effect, leads to an infrared fixed point prediction of the top quark mass. The stop sector gives important radiative contributions to the mass of the lightest Higgs boson.

\break

\begin{figure}[t]
\centering
\leavevmode
\epsfxsize=3.2in\epsffile{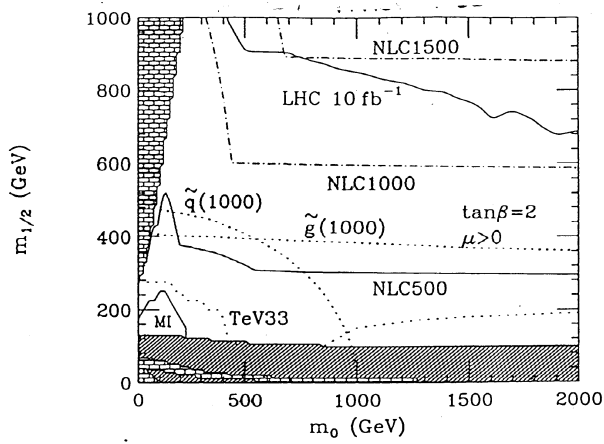}

\caption[]{Reach for supersymmetry in the mSUGRA model at various colliders (from Ref.~\cite{susy-wg}).}
\end{figure}

Search strategies for sparticles are in place for the LEP-2, Tevatron, LHC, NLC and FMC colliders. Experiments at higher energies will soon reveal whether a weak scale supersymmetry exists. A SUSY revolution would rival the excitement of the last three decades when the quark structure of matter was uncovered and the SM put in place. 

\break

\section*{Acknowlegments}

I am grateful to Chung Kao for helpful comments in the preparation of this report. This research was supported in part by the U.S.~Department of Energy under Grant No.~DE-FG02-95ER40896 and in part by the University of Wisconsin Research Committee with funds granted by the Wisconsin Alumni Research Foundation.

\end{document}